\documentclass[sigconf]{acmart}
\pdfoutput=1
\usepackage[english]{babel}
\usepackage{blindtext}
\usepackage{enumitem} 
\usepackage{algorithm}
\usepackage{subfigure}
\usepackage[noend]{algpseudocode}
\usepackage{balance}
\renewcommand\footnotetextcopyrightpermission[1]{} 
\setcopyright{none}
\newcommand{\parabf}[1]{\bigskip\noindent\textbf{#1}}

\newcommand{\cut}[1]{}
\newcommand{\etal}{{\em et al.}}

\newcommand{\sysname}{NeuroCuts\xspace}

\acmDOI{}

\acmISBN{}

\acmPrice{}

\begin{document}
\title{Neural Packet Classification}


\affiliation{%
  \institution{Eric Liang$^1$, Hang Zhu$^2$, Xin Jin$^2$, Ion Stoica$^1$}
  \institution{$^1$UC Berkeley, $^2$Johns Hopkins University}
  \institution{ekl@berkeley.edu, hzhu@jhu.edu, xinjin@cs.jhu.edu, istoica@cs.berkeley.edu}
}

\renewcommand{\shortauthors}{X. et al.}

\begin{abstract}
Packet classification is a fundamental problem in computer networking. This problem exposes a hard tradeoff between the computation and state complexity, which makes it particularly challenging. To navigate this tradeoff, existing solutions rely on complex hand-tuned heuristics, which are brittle and hard to optimize.

In this paper, we propose a deep reinforcement learning (RL) approach to solve the packet classification problem. There are several characteristics that make this problem a good fit for Deep RL. First, many existing solutions iteratively build a decision tree by splitting nodes in the tree. Second, the effects of these actions (e.g., splitting nodes) can only be evaluated once the entire tree is built. These two characteristics are naturally captured by the ability of RL to take actions that have sparse and delayed rewards. Third, it is computationally efficient to generate data traces and evaluate decision trees, which alleviate the notoriously high sample complexity problem of Deep RL algorithms.
Our solution, NeuroCuts, uses succinct representations to encode state and action space, and efficiently explore candidate decision trees to optimize for a global objective. 
It produces compact decision trees optimized for a specific set of rules and a given performance metric, such as classification time, memory footprint, or a combination of the two. 
Evaluation on ClassBench shows that NeuroCuts outperforms existing hand-crafted algorithms in classification time by 18\% at the median, and reduces both classification time and memory footprint by up to 3$\times$.
\end{abstract}

\maketitle
\section{Introduction}
\label{sec:introduction}

Packet classification is one of the fundamental problems in computer networking. The goal of packet classification is to match a given packet to a rule from a set of rules, and to do so while optimizing the classification time and/or memory footprint. Packet classification is a key building block for many network functionalities, including firewalls, access control, traffic engineering, and network measurements~\cite{HiCuts, EffiCuts, cutsplit}. As such, packet classifiers are widely deployed by enterprises, cloud providers, ISPs, and IXPs~\cite{core-router, cutsplit, tcam}.

Existing solutions for packet classification can be divided into two broad categories. Solutions in the first category are hardware-based. They leverage Ternary Content-Addressable Memories (TCAMs) to store all rules in an associative memory, and then match a packet to all these rules in parallel~\cite{tcam-sigcomm15}. As a result, TCAMs provide constant classification time, but come with significant limitations. TCAMs are inherently complex, and this complexity leads to high cost and power consumption. This makes TCAM-based solutions prohibitive for implementing large classifiers~\cite{EffiCuts}.

The solutions in the second category are software based.
These solutions build sophisticated in-memory data structures---typically decision trees---to efficiently perform packet classification~\cite{cutsplit}. While these solutions are far more scalable than TCAM-based solutions, they are slower, as the classification operation needs to traverse the decision tree from the root to the matching leaf.

Building efficient decision trees is difficult. Over the past two decades, researchers have proposed a large number of decision tree based solutions for packet classification~\cite{HiCuts,HyperCuts,hypersplit,EffiCuts,cutsplit}. However, despite the many years of research, these solutions have two major limitations.
First, they rely on hand-tuned heuristics to build the tree. Examples include maximizing split entropy~\cite{HiCuts}, balancing splits with custom space measures~\cite{HiCuts}, special handling for wildcard rules~\cite{HyperCuts}, and so on.
This makes them hard to understand and optimize over different sets of rules. If a heuristic is too general, it cannot take advantage of the characteristics of a particular set of rules. If a heuristic is designed for a specific set of rules, it typically does not achieve good results on another set of rules with different characteristics.

Second, these heuristics do not explicitly optimize for a given objective (e.g., tree depth).
They make decisions based on information (e.g., the difference between the number of rules in the children, the number of distinct ranges in each dimension) that is only \emph{loosely} related to the global objective. As such, their performance can be far from optimal.


In this paper, we propose a learning approach to packet classification. Our approach has the potential to address the limitations of the existing hand-tuned heuristics. In particular, our approach~\emph{learns} to optimize packet classification for a given set of rules and objective, can easily incorporate pre-engineered heuristics to leverage their domain knowledge, and does so with little human involvement. 
The recent successes of deep learning in solving notoriously hard problems, such as image recognition~\cite{imagenet} and language translation~\cite{translation}, have inspired many practitioners and researchers to apply deep learning, in particular, and machine learning, in general, to systems and networking problems~\cite{ddl, nas, pensieve, drl-route, schedule-drl, drl-cc, drl-pcc, verus, resource-drl}. While in some of these cases there are legitimate concerns about whether machine learning is the right solution for the problem at hand, we believe that deep learning is a good fit for our problem. This is notable since, when an efficient formulation is found, learning-based solutions have often outperformed hand-crafted alternatives~\cite{dqn-atari, alphago, learn-index}.

There are two general approaches to apply learning to packet classification. The first is to replace the decision tree with a neural network, which given a packet will output the rule matching that packet. Unfortunately, while appealing, this end-to-end solution has a major drawback: it does not guarantee the correct rule is always matched. While this might be acceptable for some applications such as traffic engineering, it is not acceptable for others, such as access control. Another issue is that large rule sets will require correspondingly large neural network models, which can be expensive to evaluate without accelerators such as GPUs. The second approach, and the one we take in this paper, is to use deep learning to build a decision tree. Recent work has applied deep learning to optimize decision trees for machine learning problems~\cite{norouzi2015efficient, xiong2017learning, Kontschieder_2015_ICCV}. These solutions, however, are designed for machine learning settings that are different than packet classification, and aim to maximize accuracy. In contrast, decision trees for packet classification provide perfect accuracy by construction, and the goal is to minimize classification time and memory footprint.


Our solution uses deep reinforcement learning (RL) to build efficient decision trees. There are three characteristics that makes RL a particularly good fit for packet classification. First, the natural solution to build a decision tree is to start with one node and recursively split (cut) it. Unfortunately, this kind of approach does not have a greedy solution. When making a decision to cut a node, we do not know whether that decision was a good one (i.e., whether it leads to an efficient tree) before we finish building the actual tree. RL naturally captures this characteristic as it does not assume that the impact of a given decision on the performance objective is known immediately. Second, unlike existing heuristics which take actions that are only loosely related to the performance objective, the explicit goal of an RL algorithm is to directly maximize the performance objective. Third, unlike other RL domains such as as robotics, for our problem it is possible to evaluate an RL model quickly (i.e., a few seconds of CPU time). This alleviates one of the main drawbacks of RL algorithms: the non-trivial learning time due to the need to evaluate a large number of models to find a good solution. By being able to evaluate each model quickly (and, as we will see, in parallel) we significantly reduce the learning time.


To this end, we design \sysname, a deep RL solution for packet classification that learns to build efficient decision trees. There are three technical challenges to formulate this problem as an RL problem. First, the tree is growing during the execution of the algorithm, as existing nodes are split. This makes it very difficult to encode the decision tree, as RL algorithms require a fixed size input. We address this problem by noting that the decision of how to split a node in the tree depends only on the node itself; it does not depend on the rest of the tree. As such, we do not need to encode the entire tree; we only need to encode the current node. The second challenge is in computing dense rewards to accelerate the learning process; here we exploit the branching structure of the problem to provide denser feedback for tree size and depth. The final challenge is that training for very large sets of rules can take a long time. To address this, we leverage RLlib~\cite{rllib}, a distributed RL library.


In summary, we make the following contributions.
\begin{itemize}[leftmargin=*]
  \item We show that the packet classification problem is a good fit for reinforcement learning (RL).

  \item We present \sysname, a deep RL solution for
  packet classification that learns to build efficient decision trees.
  

  \item We show that \sysname outperforms state-of-the-art solutions, improving classification time by 18\% at the median and reducing both time and memory usage by up to 3$\times$.
\end{itemize}

The code for \sysname is open source and is available at: \texttt{https://github.com/xinjin/neurocuts-code}  

\section{Background}
\label{sec:background}

\begin{figure}[t]
\centering
    \includegraphics[width=\linewidth]{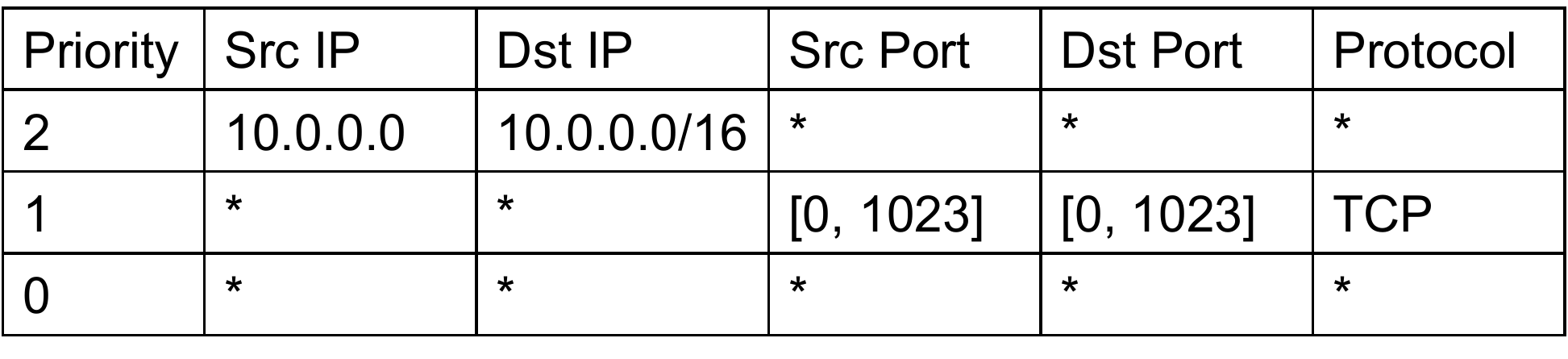}
\vspace{-0.3in}
\caption{A packet classifier example. Real-world classifiers can have 100K rules or more.}
\vspace{-0.1in}
\label{fig:background_classifier}
\end{figure}

In this section, we provide background on the packet classification problem, and summarize the key ideas behind the decision tree based solutions to solve this problem.

\subsection{Packet Classification}

A packet classifier contains a list of rules. Each rule specifies a pattern on multiple fields in the packet header. Typically, these fields include source and destination IP addresses, source and destination port numbers, and protocol type. The rule's pattern specifies which packets match the rule. Matching conditions include prefix based matching (e.g., for IP addresses), range based matching (e.g., for port numbers), and exact matching (e.g., for protocol type). A packet matches a rule if each field in the packet header satisfies the matching condition of the corresponding field in the rule, e.g., the packet's source/destination IP address matches the prefix of the source/destination address in the rule, the packet's source/destination port number is contained in the source/destination range specified in the rule, and the packet's protocol type matches the rule's protocol type.   

Figure~\ref{fig:background_classifier} shows a packet classifier with three rules. The first rule matches all packets with source address 10.0.0.1 and the destination addresses sharing prefix 10.0.0.0/16. Other fields are unspecified (i.e., they are $\star$) meaning that the rule matches any value in these fields. The second rule matches all TCP packets with source and destination ports in the range [0, 1023], irrespective of IP addresses (as they are $\star$). Finally, the third rule is a default rule that matches all packets. This guarantees that any packet matches at least one rule.

Since rules can overlap, it is possible for a packet to match multiple rules. To resolve this ambiguity, each rule is assigned a priority. A packet is then matched to the highest priority rule. For example, packet  (10.0.0.0, 10.0.0.1, 0, 0, 6) matches all the three rules of the packet classifier in Figure~\ref{fig:background_classifier}. However, since the first rule has the highest priority, we match the packet to the first rule only.

\subsection{Decision Tree Algorithms}


Packet classification is similar to the point location problem in a multi-dimensional geometric space: the fields in the packet header we are doing classification on (e.g., source and destination IP addresses, source and destination port numbers, and protocol number) represent the dimensions in the geometric space, a packet is represented as a point in this space, and a rule as a hypercube. Unfortunately, the point location problem exhibits a hard tradeoff between time and space complexities~\cite{algorithms-pc}.

The packet classification problem is then equivalent to finding all hypercubes that contains the point corresponding to a given packet.
 In particular, in a $d$-dimensional geometric space with $n$ non-overlapping hypercubes and when $d>3$, this problem has either $(i)$ a lower bound of $O(log~n)$ time and $O(n^d)$ space, or $(ii)$ a lower bound of $O(log^{d-1} n)$ time and $O(n)$ space~\cite{algorithms-pc}. The packet classification problem allows the hypercubes (i.e., rules) to overlap, and thus is at least as hard as the point location problem~\cite{algorithms-pc}. In other words, if we want logarithmic computation time, we need space that is exponential in the number of dimensions (fields), and if we want linear space, the computation time will be exponential in the logarithm of the number of rules. Given that for packet classification $d = 5$, neither of these choices is attractive. 

Next, we discuss two common techniques employed by existing solutions to build decision trees for packet classification: \emph{node cutting} and \emph{rule partition}.

\vskip 0.1in
\noindent
{\bf Node cutting.}
Most existing solutions for packet classification aim to build a decision tree that exhibits low classification time (i.e., time complexity) and memory footprint (i.e., space complexity)~\cite{EffiCuts}. The main idea is to split nodes in the decision tree by ``cutting'' them along one or more dimensions. Starting from the root which contains all rules, these algorithms iteratively split/cut the nodes until each leaf contains fewer than a predefined number of rules.
Given a decision tree, classifying a packet reduces to walk the tree from the root to a leaf, and then chose the highest priority rule associated with that leaf.

\begin{figure}[t]
\centering
    \includegraphics[width=\linewidth]{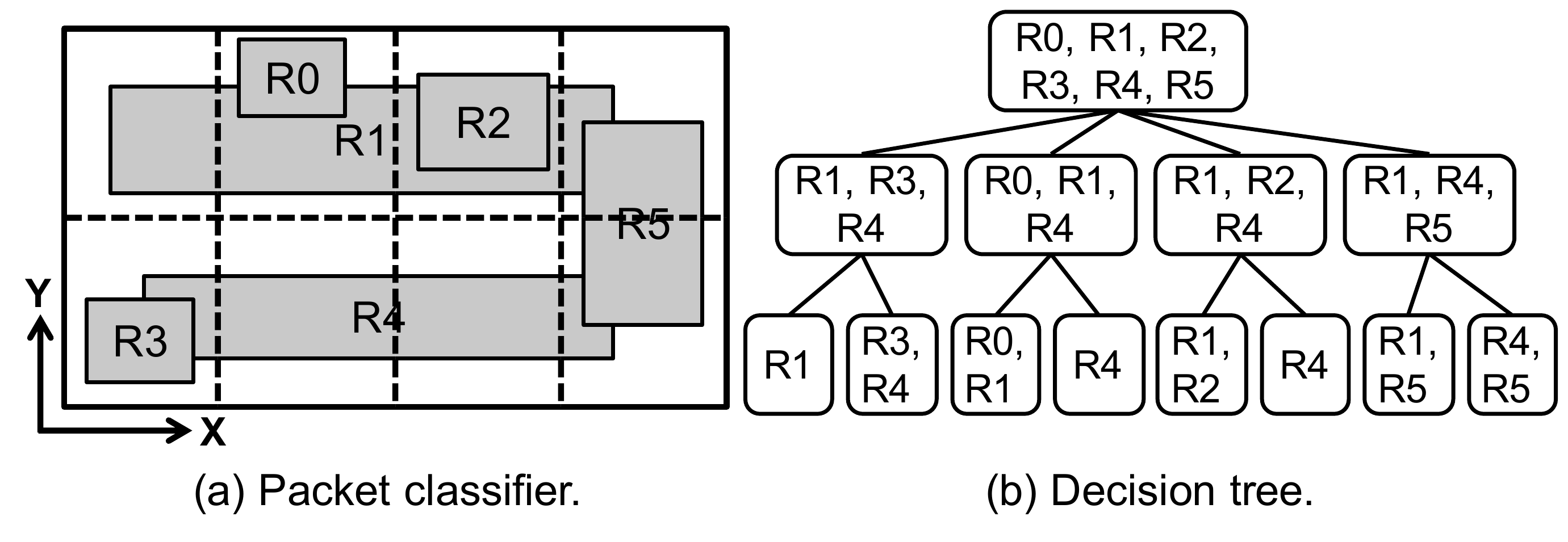}
\vspace{-0.3in}
\caption{Node cutting.}
\vspace{-0.05in}
\label{fig:background_cut}
\end{figure}

\begin{figure}[t]
\centering
    \includegraphics[width=\linewidth]{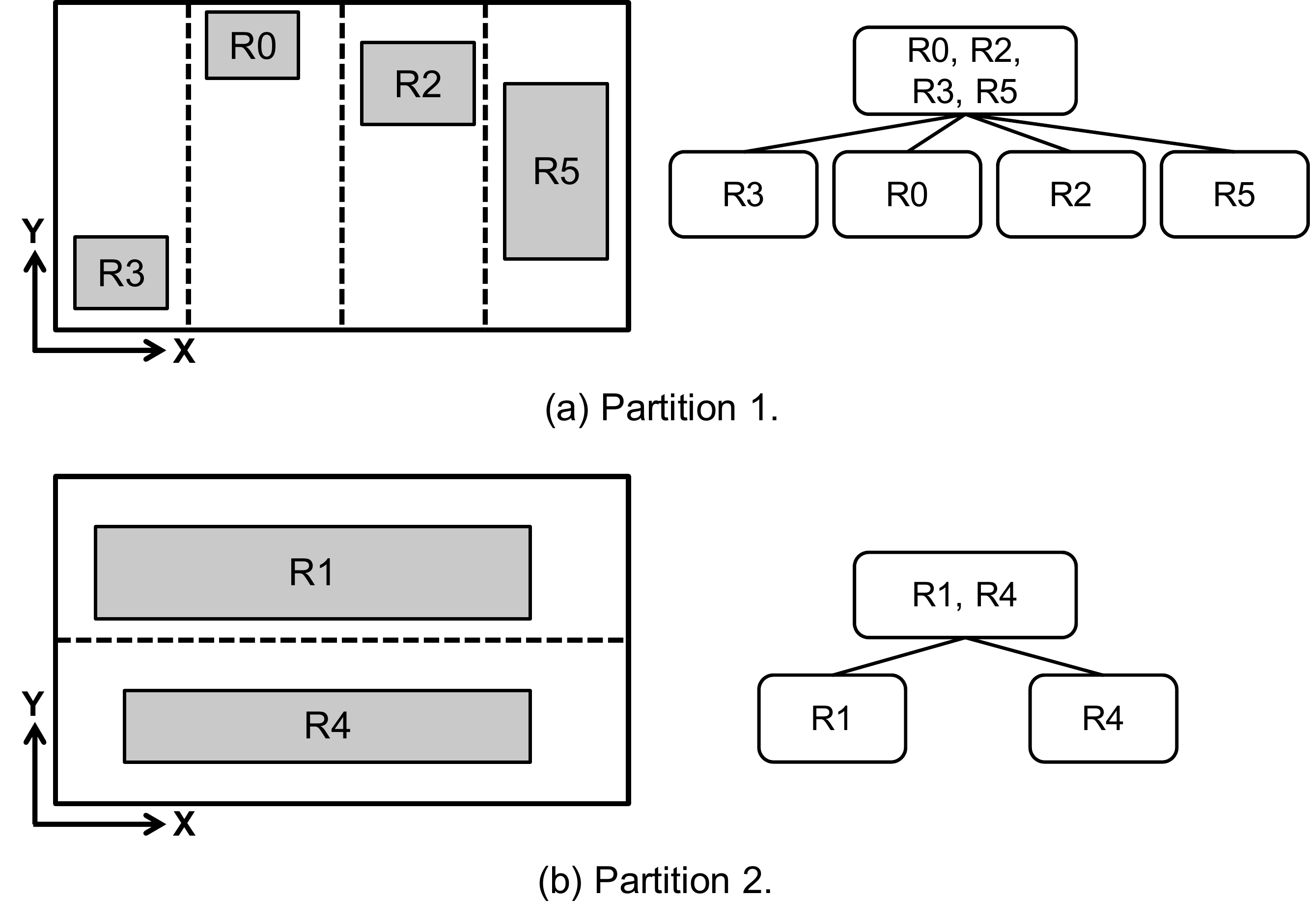}
\vspace{-0.3in}
\caption{Rule partition.}
\vspace{-0.1in}
\label{fig:background_partition}
\end{figure}

Figure~\ref{fig:background_cut} illustrates this technique. The packet classifier contains six rules (R0 to R5) in a two-dimensional space. Figure~\ref{fig:background_cut}(a) shows each rule as a rectangle in the space, and represents the cuts as dashed lines. Figure~\ref{fig:background_cut}(b) shows the corresponding decision tree for this packet classifier. The root of the tree contains all the six rules. First, we cut the entire space (which represents the root) into four chunks along dimension $x$. This leads to the creation of four children. If a rule intersects a child's chunk, it is added to that child. For example, R1, R3 and R4 all intersect the first chunk (i.e., the first quarter in this space), and thus they are all added to the first root's child. If a rule intersects multiple chunks it is added to each corresponding child, e.g., R1 is added to all the four children. Next, we cut the chunk corresponding to each of the four children along dimension $y$. As a result, each of the nodes at the first level will end up with two children.

\vskip 0.1in
\noindent
{\bf Rule partition.}
One challenge with "blindly" cutting a node is that we might end up with a rule being replicated to a large number of nodes~\cite{EffiCuts}. In particular, if a rule has a large size along one dimension, cutting along that dimension will result in that rule being added to many nodes. For example, rule R1 in Figure~\ref{fig:background_cut}(a) has a large size in dimension $x$. Thus, when cutting along dimension $x$, R1 will end up being replicated at every node created by the cut. Rule replication can lead to decision trees with larger depths and sizes, which translate to higher classification time and memory footprint.  

One solution to address this challenge is to first partition rules based on their "shapes". Broadly speaking, rules with large sizes in a particular dimension are put in the same set. Then, we can build a separate decision tree for each of these partitions.
Figure~\ref{fig:background_partition} illustrates this technique. The six rules in Figure~\ref{fig:background_cut} are grouped into two partitions. One partition consists of rules R1 and R4, as both these rules have large sizes in dimension $x$. The other partition consists of the other four rules, as these rules have small sizes in dimension $x$. Figure~\ref{fig:background_partition}(a) and Figure~\ref{fig:background_partition}(b) show the corresponding decision trees for each partition. Note that the resulting trees have lower depth, and smaller number of rules per node as compared to the original decision tree in Figure~\ref{fig:background_cut}(b). To classify a packet, we classify it against every decision tree, and then choose the highest priority rule among all rules the packet matches in all decision trees.

\vskip 0.1in
\noindent
{\bf Summary.} Existing solutions build decision trees by employing two types of actions: node cutting and rule partition. These solutions mainly differ in the way they decide (i) at which node to apply the action, (ii) which action to apply, and (iii) how to apply it (e.g., along which dimension(s) to partition). 



\section{A Learning-Based Approach}
\label{sec:motivation}

In this section, we describe a learning-based approach for packet classification. We motivate our approach, discuss the formulation of classification as a learning problem, and then present our solution.

\subsection{Why Learn?}

The existing solutions for packet classification rely on hand-tuned heuristics to build decision trees. Unfortunately, this leads to two major limitations.

First, these heuristics often face a difficult~\emph{trade-off} between~\emph{performance} and~\emph{cost}. Tuning such a heuristic for a given set of rules is an expensive proposition, requiring considerable human efforts and expertise. Worse yet, when given a different rule set, one might have to do this all over again. Addressing this  challenge has been the main driver of a long line of research over the past two decades~\cite{HiCuts,HyperCuts,hypersplit,EffiCuts,cutsplit}. Of course, one could build a general heuristic for a large variety of rule sets. Unfortunately, such a solution would not provide the best performance for a given set of rules.

Second, existing algorithms do not directly optimize for a global objective. Ideally, a good packet classification solution should optimize for $(i)$ classification time, $(ii)$ memory footprint, or $(iii)$ a combination between the two. Unfortunately, the existing heuristics do not \emph{directly} optimize for any of these objectives. At their core, these heuristics make greedy decisions to build decision trees. At every step, they decide on whether to cut a node or partition the rules based on simple statistics (e.g., the size of the rules in each dimension, number of unique ranges in each dimension), which are poorly correlated with the desired objective. As such, the resulting decision trees are often far from being optimal. 

As we will see, a learning-based approach can address these limitations. Such an approach can learn to generate an efficient decision tree for a specific set of rules without the need to rely on hand-tuned heuristics. This is not to say these heuristics do not have value; in fact they often contain key domain knowledge that we show can be leveraged and improved on by the learning algorithm.

\subsection{What to Learn?}

Classification is a central task in machine learning literature. The recent success of using deep neural networks (DNNs) for image recognition, speech recognition and language translation has been single-handedly responsible for the recent AI "revolution"~\cite{imagenet,translation,speech}. 

As such, one natural solution for packet classification would be to replace a decision tree with a DNN. In particular, such DNN will take as input the fields of a packet header and output the rule matching that packet. Related to our problem, prior work has shown that DNN models can be effectively used to replace B-Trees for indexing~\cite{learn-index}.

However, this solution has two drawbacks. First, a DNN-based classifier does not guarantee 100\% accuracy. This is because training a DNN is fundamentally a stochastic process. Second, given a DNN packet classification result, it is expensive to verify whether the result is correct or not. Unlike the recently proposed learned index solution to replace B-Trees~\cite{learn-index}, the rules in packet classification are multi-dimensional and overlap with each other. If a rule matches a packet, we still need to check other rules to see if this rule has the highest priority
among all matched rules.


To avoid these drawbacks, in this paper we propose to learn building decision trees for a given set of rules. Since the result is still a decision tree, we can guarantee correctness, and it will be easy to deploy the classifier with existing systems (hardware and software) compared to a DNN.

\begin{figure}[t]
    \centering
    \subfigure[]{
        \label{fig:overview_rl}
        \includegraphics[width=\linewidth]{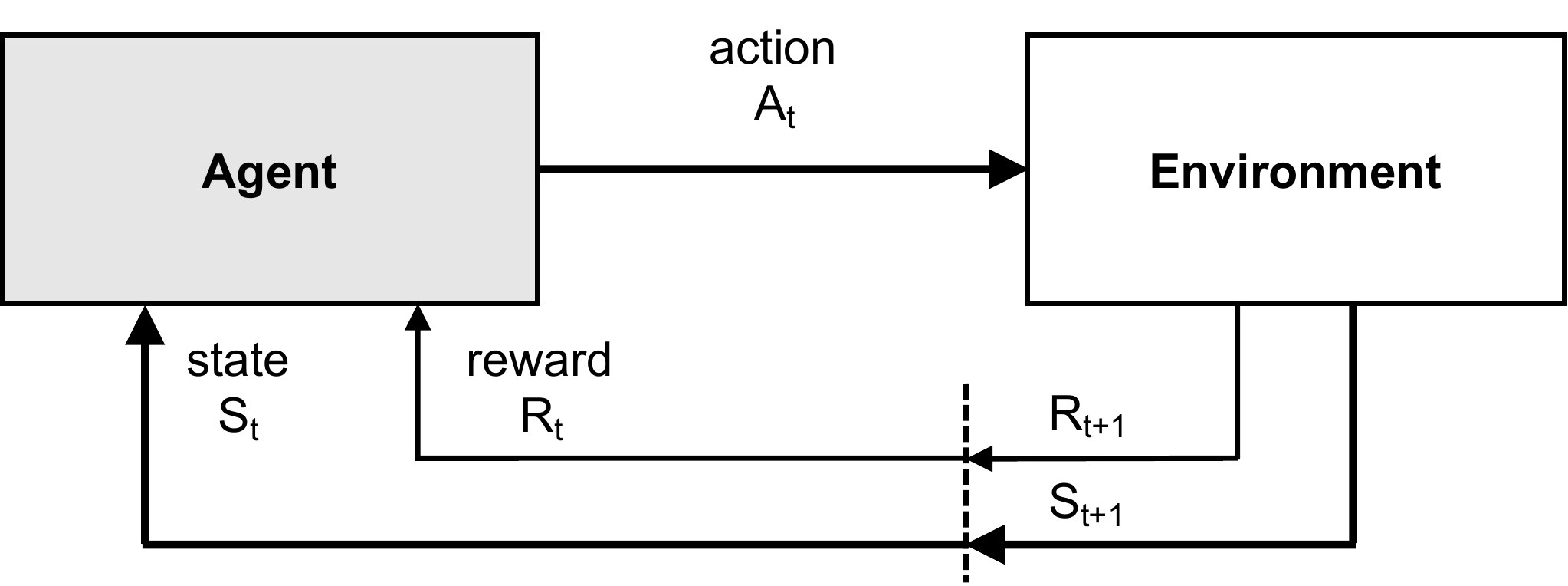}}
    \vspace{-.1in}
    \subfigure[]{
        \label{fig:overview_ar}
        \includegraphics[width=\linewidth]{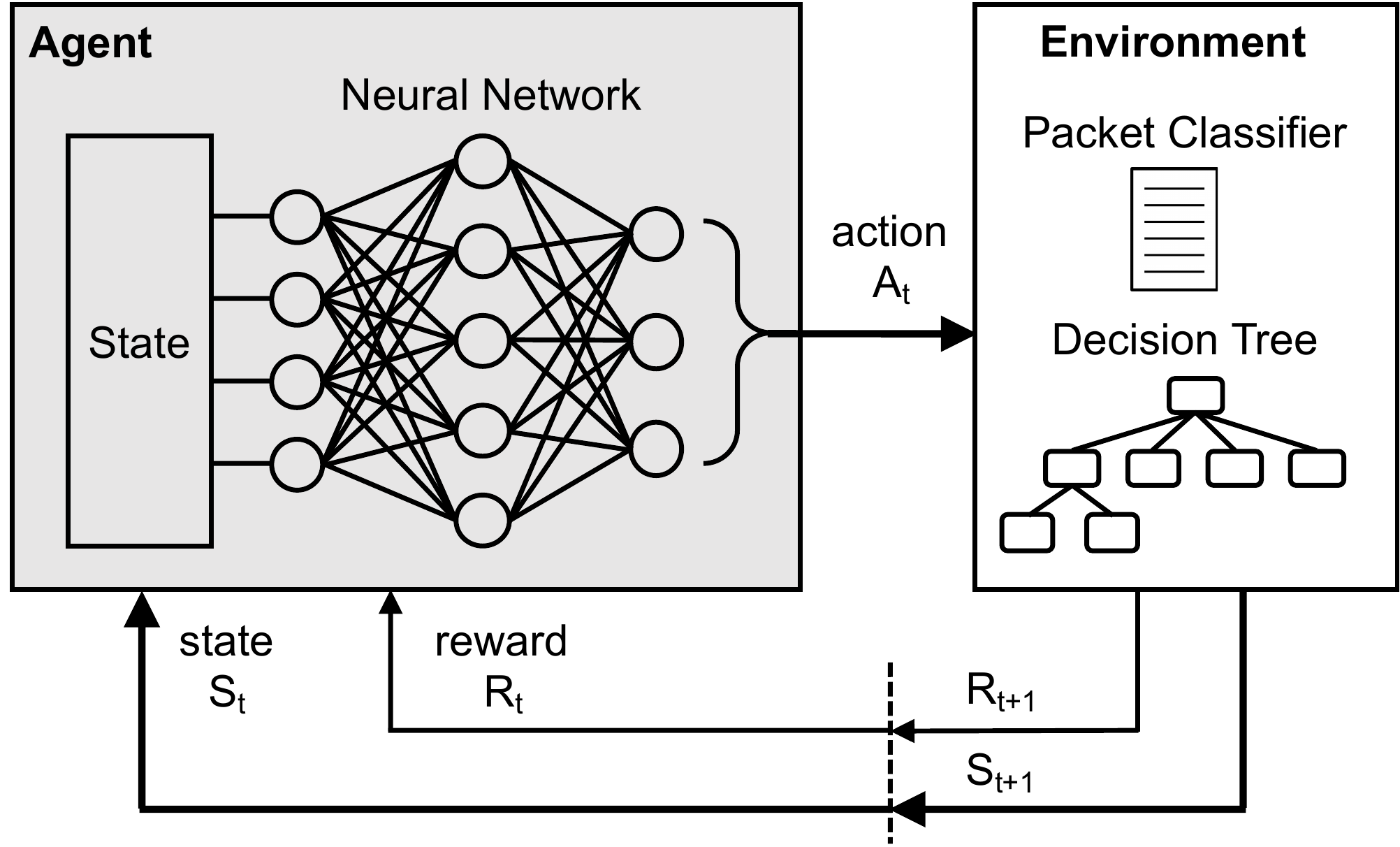}}
    \vspace{-0.1in}
    \caption{(a) Classic RL system. An agent takes an action, $A_t$, based on the current state of the environment, $S_t$, and applies it to the environment. This leads to a change in the environment state ($S_{t+1}$) and a reward ($R_{t+1}$). (b) \sysname as an RL system.}
    \vspace{-0.2in}
    \label{fig:overview_architecture}
\end{figure}

\subsection{How to Learn?}

In this section, we show that the problem of building decision trees maps naturally to RL. As illustrated in Figure~\ref{fig:overview_architecture}(a), an RL system consists of an agent that repeatedly interacts with an environment. The agent observes the state of the environment, and then takes an action that might change the environment's state. The goal of the agent is to compute a \emph{policy} that maps the environment's state to an action in order to optimize a reward. As an example, consider an agent playing chess. In this case, the environment is the board, the state is the position of the pieces on the board, an action is moving a piece on the board, and the reward could be $1$ if the game is won, and $-1$, if the game is lost. 

This simple example illustrates two characteristics of RL that are a particularly good fit to our problem. First, rewards are \emph{sparse}, i.e., not every state has associated a reward. For instance, when moving a piece we do not necessary know whether that move will result in a win or loss. Second, the rewards are \emph{delayed}; we need to wait until the end of the game to see whether the game was won or lost. 

To deal with large state and action spaces, recent RL solutions have employed DNNs to implement their policies. These solutions, called Deep RL, have achieved remarkable results matching humans at playing Atari games~\cite{dqn-atari}, and beating the Go world champion~\cite{alphagozero}. These results have encouraged researchers to apply Deep RL to networking and systems problems, from routing, to congestion control, to video streaming, and to job scheduling~\cite{ddl, nas, pensieve, drl-route, schedule-drl, drl-cc, drl-pcc, verus, resource-drl}.
Building a decision tree can be easily cast as an RL problem: the environment's state is the current decision tree, an action is either cutting a node or partitioning a set of rules, and the reward is either the classification time, memory footprint, or a combination of the two.
While in some cases there are legitimate concerns about whether Deep RL is the right solution for the problem at hand, we identify several characteristics that make packet classification a particularly good fit for Deep RL.

First, when we take an action, we do not know for sure whether it will lead to a good decision tree or not; we only know this once the tree is built. As a result, the rewards in our problem are both {\em sparse} and {\em delayed}. This is naturally captured by the RL formulation.

Second, the explicit goal of RL is to maximize the reward. Thus, unlike existing heuristics, our RL solution aims to explicitly optimize the performance objective, rather than using local statistics whose correlation to the performance objective can be tenuous. 

Third, one potential concern with Deep RL algorithms is sample complexity. In general, these algorithms require a huge number of samples (i.e., input examples) to learn a good policy. Fortunately, in the case of packet classification we can generate such samples cheaply. A sample, or rollout, is a sequence of actions that builds a decision tree with the associated reward(s) by using a given policy. The reason we can generate these rollouts cheaply is because we can build all these trees in software, and do so in parallel. Contrast this with other RL-domains, such as robotics, where generating each rollout can take a long time and requires expensive equipment (i.e., robots).

\section{\sysname Design}
\label{sec:design}

\subsection{\sysname Overview}

We introduce the design for \sysname, a new Deep RL formulation of the packet classification problem.
Given a rule set and an objective function (i.e., classification time, memory footprint, or a combination of both), \sysname learns to build a
decision tree that minimizes the objective.

\begin{figure*}
  \centering
    \centering
  \includegraphics[width=10cm,clip,trim={0cm 4.9cm 0cm 0cm}]{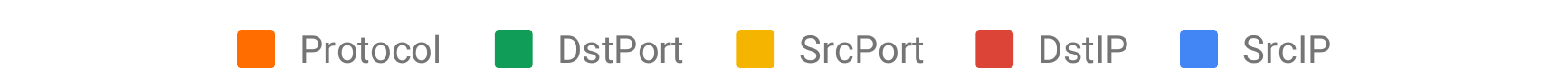}
  \begin{subfigure}[NeuroCuts starts with a randomly initialized policy that generates poorly shaped trees (left, truncated). Over time, it learns to reduce the tree depth and develops a more coherent strategy (center). The policy converges to a compact depth-12 tree (right) that specializes in cutting \texttt{SrcIP}, \texttt{SrcPort}, and \texttt{DstPort}.]{
      \includegraphics[width=3.8cm]{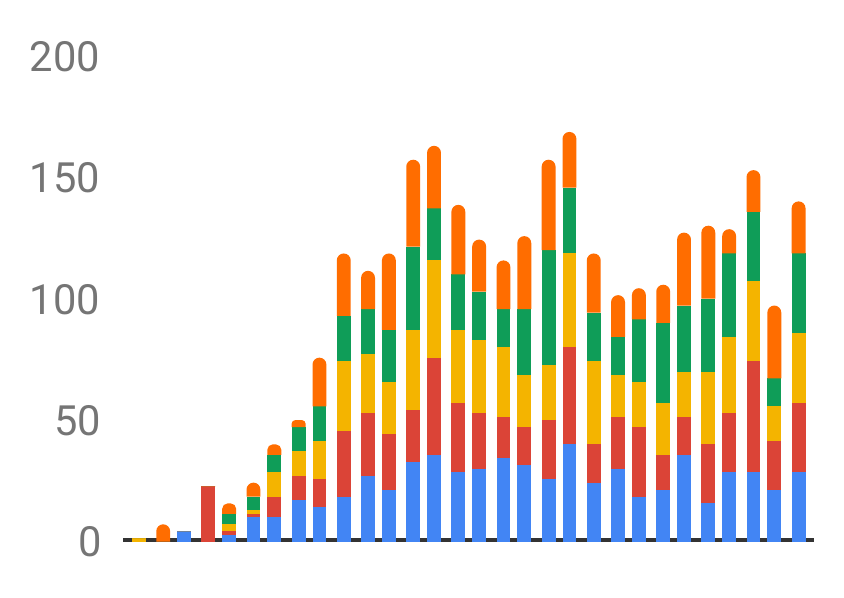}    
      \includegraphics[width=0.6cm]{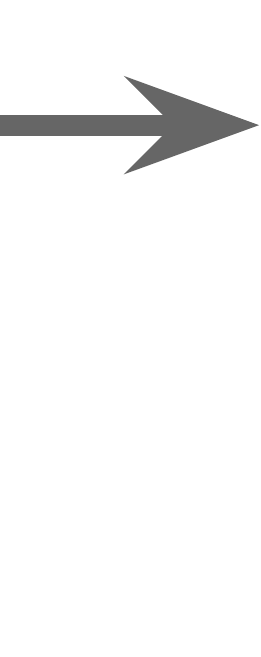}  
      \includegraphics[width=3.8cm]{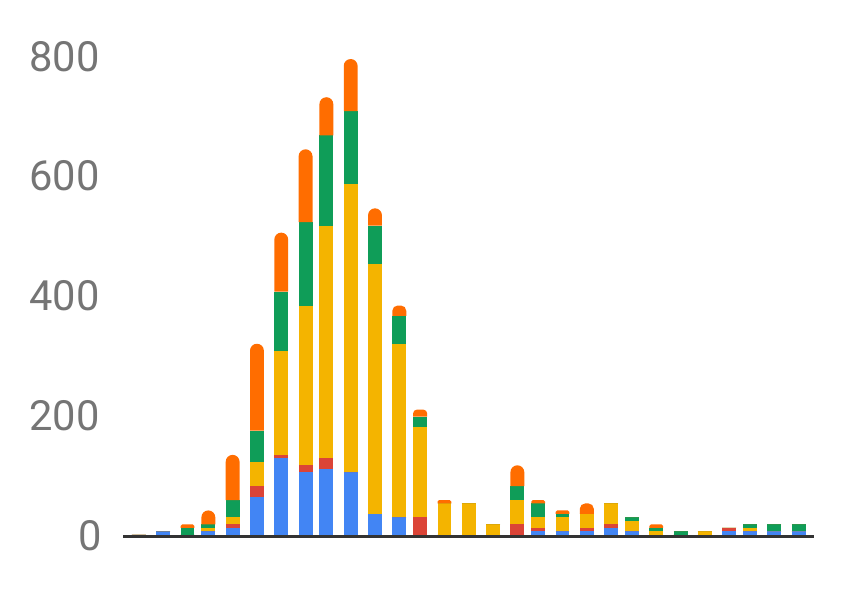}
      \includegraphics[width=0.6cm]{figures/arrow.pdf}  
      \includegraphics[width=3.8cm]{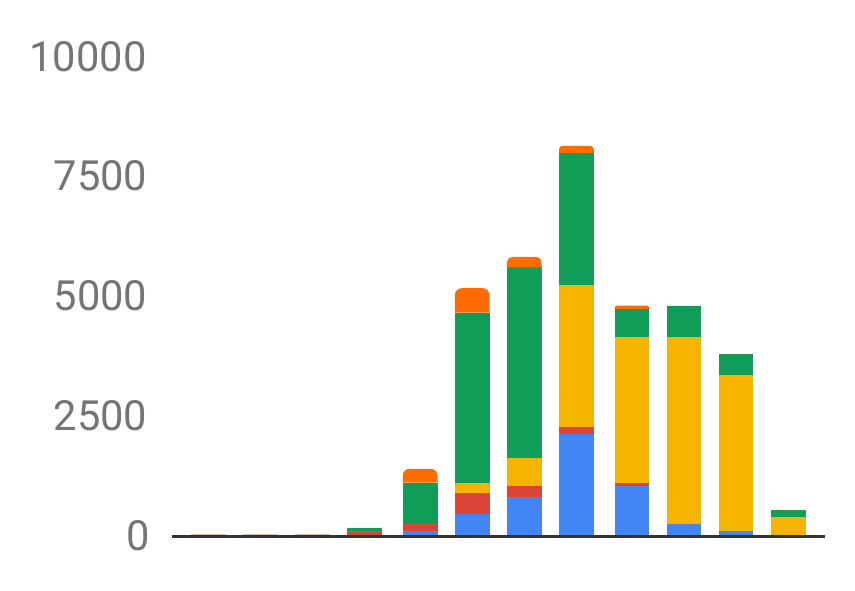}
  }
  \end{subfigure}
  \hspace{.3cm}
  \begin{subfigure}[In comparison, HiCuts produces a depth-29 tree for this rule set that is 15$\times$ larger and 3$\times$ slower in classification time.]{
      \includegraphics[width=3.8cm]{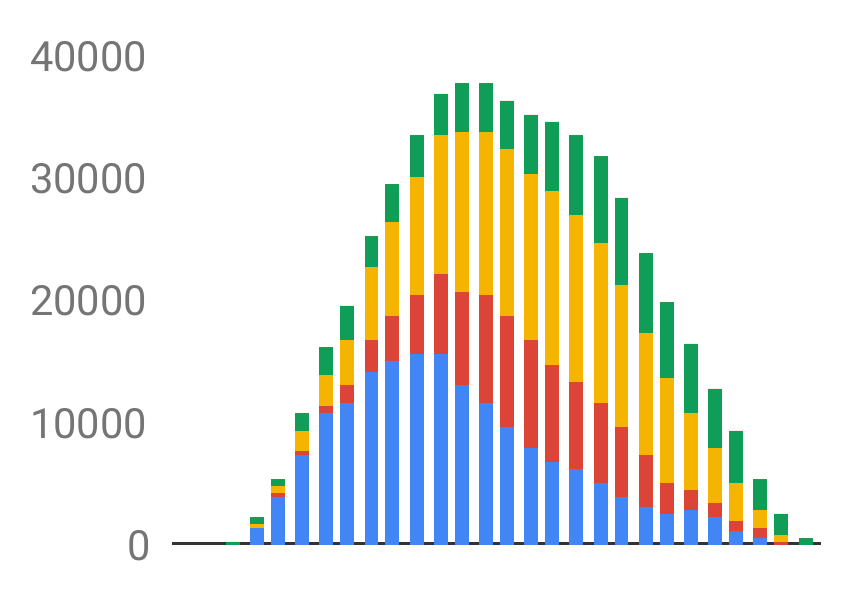}
  }
  \end{subfigure}
   \vspace{-.3cm}
    \caption{Visualization of NeuroCuts learning to split the \textbf{fw5\_1k} ClassBench rule set. The x-axis denotes the tree level, and the y-axis the number of nodes at the level. The distribution of cut dimensions per level of the tree is shown in color.}
       \vspace{-.1cm}
    \label{fig:training_vis}
\end{figure*}

Figure~\ref{fig:overview_architecture}(b) illustrates the framing of \sysname as an RL system: the environment consists of the set of rules and the current decision tree, while the \emph{agent} uses a model (implemented by a DNN) that aims to select the best cut or partition action to incrementally build the tree. A cut action divides a node along a chosen dimension (i.e., one of \texttt{SrcIP}, \texttt{DstIP}, \texttt{SrcPort}, \texttt{DstPort}, and \texttt{Protocol}) into a number of sub-ranges (i.e., 2, 4, 8, 16, or 32 ranges), and creates that many child nodes in the tree. A partition action on the other hand divides the rules of a node into disjoint subsets (e.g., based on the coverage fraction of a dimension), and creates a new child node for each subset.
The available actions for the current node are advertised by the environment at each step, the agent chooses among them to generate the tree, and over time the agent learns to optimize its decisions to maximize the reward from the environment.
Figure~\ref{fig:training_vis} visualizes the learning process of \sysname.


\subsection{\sysname Training Algorithm}

Recall that the goal of an RL algorithm is to compute a policy to maximize rewards from the environment. Referring again to Figure~\ref{fig:overview_architecture}, the environment defines the action space $A$ and state space $S$. The agent starts with an initial policy, evaluates it using multiple rollouts, and then updates it based on the results (rewards) of these rollouts.  Then, it repeats this process until satisfied with the reward.

We first consider a strawman formulation of decision tree generation as a single Markov Decision Process (MDP). In this framing, a rollout begins with a tree consisting of a single node. This is the initial state, $s_0 \in S$. At each step $t$, the agent executes an action $a_t \in A$ and receives a reward $r_t$; the environment transitions from the current state $s_t \in S$ to the next state $s_{t+1} \in S$ (i.e., the updated tree and next node to process). The goal is to maximize the total reward received by the agent, i.e., $\sum_t \gamma^t r_t$ where $\gamma$ is a discounting factor used to prioritize more recent rewards.


\vskip 0.1in
\noindent
{\bf Design challenges.}
While at a high level this RL formulation seems straightforward, there are three key challenges we need to address before we have a realizable implementation. The first is how to encode the variable-length decision tree state $s_t$ as an input to the neural network policy.
While it is possible to flatten the tree, say, into an 1-dimensional vector, the size of such a vector would be very large (i.e., hundreds of thousands of units). This will require both a very large network model to process such input, and a prohibitively large number of samples. 


While recent work has proposed leveraging recurrent
neural networks (RNNs) and graph embedding techniques~\cite{graph-emb, graphemb-aaai, non-graphemb} to reduce the input size, these solutions are brittle in the face of large or dynamically growing graph structures \cite{zhou2018graph}.
Rather than attempting to solve the state representation problem to deal with large inputs, in \sysname we instead take advantage of the underlying structure of packet classification trees to design a simple and compact state representation. This means that when the agent is deciding how to split a node, it only observes a fixed-length representation of the node. All needed state is encoded in the representation; no other information about the rest of the tree is observed.

The second challenge is how to deal with the sparse and delayed rewards incurred by the node-by-node process of building the decision tree. While we could in principle return a single reward to the agent when the tree is complete, it would be very difficult to train an agent in such an environment. Due to the long length of tree rollouts (i.e., many thousands of steps), learning is only practical if we can compute meaningful \textit{dense rewards}.\footnote{Note that just returning -1 or -$cutSize$ for each step would not be a particularly useful dense reward.} Such a dense reward for an action would be based on the statistics of the subtree it leads to (i.e., its depth or size).\footnote{The rewards for NeuroCuts correspond to the true problem objective; we do not do "reward engineering" since that would bias the solution.} Unfortunately, it is not possible to compute this until the subtree is complete. To handle this, we take the somewhat unusual step of only computing rewards for the rollout when the tree is completed, and setting $\gamma=0$, effectively creating a series of 1-step decision problems similar to contextual bandits \cite{bandits}. However, unlike the bandit setting, these 1-step decisions are connected through the dynamics of the tree building process.

Another way of looking at the dense reward problem is that the process of building a decision tree is not really sequential but \textit{tree-structured} (i.e., it is more accurately modeled as a branching decision process \cite{kolmogorov1947stochastic, branch1, branch2}), and we need to account for the reward calculations accordingly. In such a "branching" formulation, $\gamma >  0$, but the rewards of an action are computed as an aggregation over \textit{multiple child states} produced by an action. For example, cutting a node produces multiple child sub-nodes, and the reward calculation may involve a \texttt{sum} or a \texttt{min} over each child's future rewards, depending on whether we are optimizing for tree size or depth. The 1-step decision problem and branching decision process formulations of \sysname are roughly equivalent; in the implementation section we describe how we adapt standard RL algorithms to run \sysname.

The final challenge is how to scale the solution to large packet classifiers. The decision tree for a packet classifier with 100K rules can have hundreds of thousands of nodes. The size of the tree impedes training along several dimensions. Not only does it take more steps to finish building a tree, but the execution time of each action increases as there are more rules to process. The space of trees to explore is also larger, requiring the use of larger network models and generating more rollouts to train.
\begin{figure}
  \centering

      \includegraphics[width=2cm]{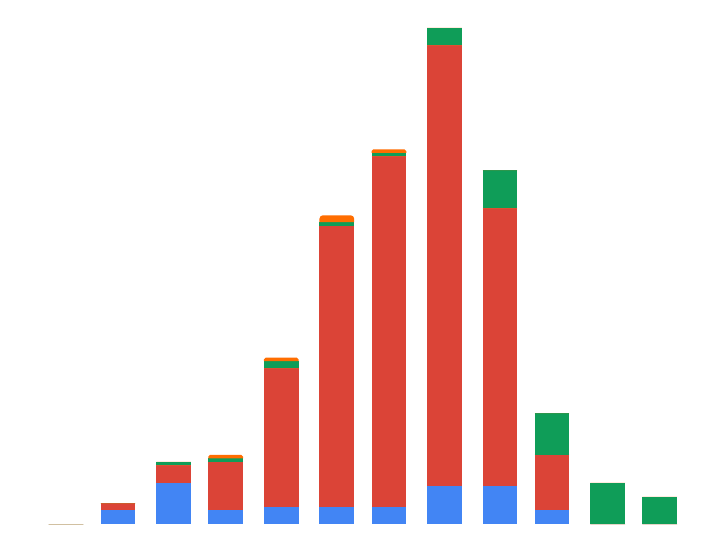}    
      \includegraphics[width=2cm]{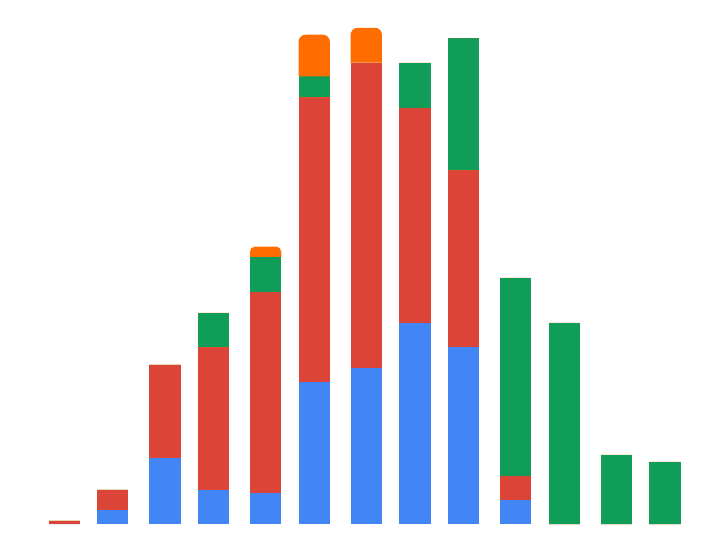}
      \includegraphics[width=2cm]{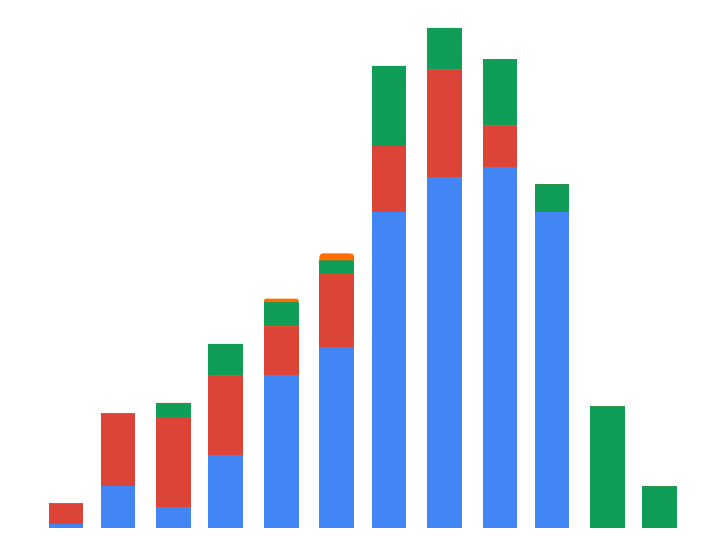}  
      \includegraphics[width=2cm]{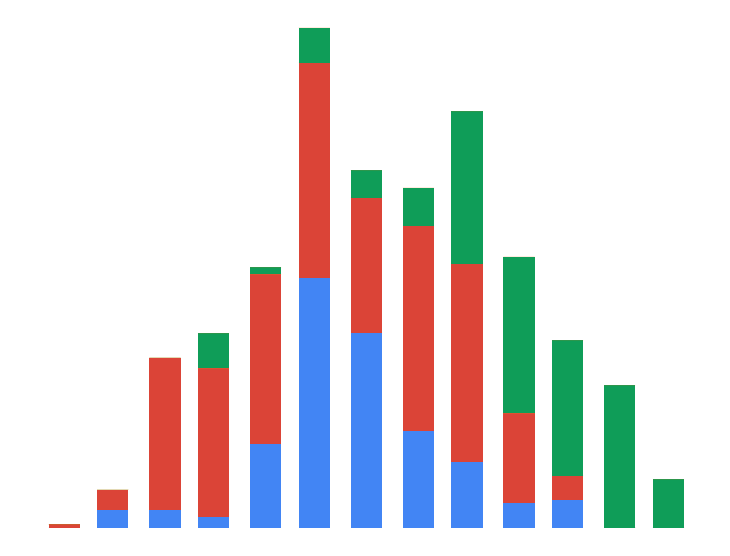}

    \caption{The NeuroCuts policy is stochastic, which enables it to effectively explore many different tree variations during training.
    Here we visualize four random tree variations drawn from a single policy trained on the \textbf{acl4\_1k} ClassBench rule set.}
    

    \label{fig:neurocuts_stochasticity}
\end{figure}

\vskip 0.1in
\noindent
{\bf State representation.} One key observation is that the action on a tree node only depends on the node itself, so it is not necessary to encode the entire decision tree in the environment state. Our goal to optimize a global performance objective over the entire tree suggests that we would need to make decisions based on the global state.  However, this does not mean that the state representation needs to encode the entire decision tree. Given a tree node, the action on that node only needs to make the best decision to optimize the sub-tree rooted at that node. It does not need to consider other tree nodes in the decision tree. 

Formally, given tree node $n$, let $t_n$ and $s_n$ denote $n$'s classification time and memory footprint, respectively, and $T_n$ and $S_n$ be the classification time and memory footprint
of the entire sub-tree rooted at node $n$, respectively. Then, for a cut action, we have the following equations:

\vskip -0.1in
\begin{align}
T_n = t_n + \mbox{max}_{i \in children(n)} T_i \\
S_n = s_n + \mbox{sum}_{i \in children(n)} S_i
\end{align}

\noindent
Similarly, for a partition action, we have 

\vskip -0.1in
\begin{align}
T_n = t_n + \mbox{sum}_{i \in children(n)} T_i \\
S_n = s_n + \mbox{sum}_{i \in children(n)} S_i
\end{align}

An action, $a$, taken on node $n$ only needs to optimize the sub-tree rooted at $n$ according to the following expression, 

\vskip -0.1in
\begin{align}
V_n = \mbox{argmax}_{a \in A} -(c \cdot T_n + (1-c) \cdot S_n),
\end{align}

\noindent
where $c$ is a coefficient capturing the tradeoff between classification time and memory footprint.
The negation is needed since we want to \textit{minimize} time and space complexities.

When $c \in \{0, 1\}$, it is easy to see that if at every tree node $n$ we take the action that optimizes $V_n$, then, by induction, we end up optimizing $V_r$, where $r$ is the root of the tree. In other words, we end up optimizing the global objective (reward) for the entire decision tree. For $0 < c < 1$ this optimization becomes approximate, but we find empirically that $c$ can still be used to interpolate between the two objectives. It is important to note here that while the state representation only encodes current node $n$, action $a$ taken for node $n$ is \emph{not local}, as it optimizes the \emph{entire sub-tree} rooted at $n$.

In summary, we only need to encode the current node as the input state of the agent. This is because the environment builds the tree node-by-node, node actions need only consider their own state, and each node contains a subset of the rules of its parent (i.e., rules contained in some subspace of its parent space). Therefore, nodes in the tree can be completely defined by the ranges they occupy in each dimension. Given $d$ dimensions, we use $2d$ numbers to encode a tree node, which indicate the left and right boundaries of each dimension for this node. The state also needs to describe the partitioning at the node, which can be handled in a similar way. We defer a full description of the NeuroCuts state and action representations to Appendix \ref{appendix:spaces}.

\vskip 0.1in
\noindent
{\bf Training algorithm.} We use an actor-critic algorithm to train the agent's policy~\cite{actorcritic-alg}.
This class of algorithms have been shown to provide state-of-the-art results in many use cases~\cite{autoaugment, a3c, ppo}, and can be easily scaled to the distributed setting~\cite{espeholt2018impala}. We also experimented with Q-learning \cite{dqn} based approaches, but found they did not perform as well. 


Algorithm~\ref{alg:training} shows the pseudocode of the \sysname algorithm, which executes as follows. \sysname starts with the root node of the decision tree, $s^*$. The end goal is to learn an optimized stochastic policy function
$\pi(a|\mathbf{s}; \theta)$ (i.e., the actor). \sysname first initializes
all the parameters (line 1-6), and then runs for $N$ rollouts to train the policy and the value function (line 7-23). After each rollout, it reinitializes the decision tree to the root node
(line 9). It then incrementally builds the tree by repeatedly selecting and applying an
action on each non-terminal leaf node (line 11-13) according to the current policy. A terminal leaf node is a node in which the number of rules is below a given threshold. 

More specifically, \sysname traverses the tree nodes in depth-first-search (DFS) order (line 13), i.e., it recursively cuts the child of the current node until the node becomes a terminal leaf. Note that the DFS order is not essential. It is used to give a way for the agent to find a tree node to cut. Other orders, such as the breadth-first-search (BFS), can be used as well. After the decision tree is built, the gradients are reset (line 14), and then the algorithm iterates over all the tree nodes to aggregate the gradients (line 15-21). Finally, \sysname uses the gradients to update the parameters of the actor and critic networks (line 22), and proceeds to the next rollout (line 23).

The first gradient computation (line 19) corresponds to that for the \textit{policy gradient loss}. This loss defines the direction to update $\theta$ to improve the expected reward. An estimation of the state value $V(s; \theta_v)$ is subtracted from the rollout reward $R$ to reduce the gradient variance \cite{konda2000actor}. $V$ is trained concurrently to minimize its prediction error (line 21).
Figure~\ref{fig:training_vis} visualizes the learning process of \sysname to build a decision tree.
The NeuroCuts policy is stochastic, enabling it to effectively explore many different tree variations during training, as illustrated in Figure~\ref{fig:neurocuts_stochasticity}.

\begin{algorithm}[t]
\footnotesize
\caption{Learning a tree-generation policy using an actor-critic algorithm.}
\begin{flushleft}
\textbf{Input:} The root node $\mathbf{s}^*$ where a tree always grows from. \\
\vspace{0.1in}
\textbf{Output:} A stochastic policy function $\pi(a|\mathbf{s}; \theta)$ that outputs a
branching action $a \in \mathcal{A}$ given a node state $\mathbf{s}$, and a
value function $V(\mathbf{s}; \theta_v)$ that outputs a value estimate for a
node state.
\vspace{0.1in}

\textbf{Main routine:}
\begin{algorithmic}[1]
\State \textbf{// Initialization}

\State Randomly initialize the model parameters $\theta$, $\theta_v$
\State Maximum number of rollouts $N$
\State Coefficient $c \in [0, 1]$ that trades off classification time vs. space
\State Reward scaling function $f(x) \in \{x, \textsc{log}(x)\}$
\State $n \gets 0$\;

\State \textbf{// Training}
\While{$n < N$}
    \State $s \gets \textsc{Reset}(\mathbf{s^*})$
    \State \textbf{// Build a tree using the current policy}
    \While{$s \ne \textsc{Null}$}
        \State $a \gets \pi(a|s; \theta)$
        \State $s \gets \textsc{GrowTreeDFS}(s, a)$
    \EndWhile
    \State Reset gradients $d\theta \gets 0$ and $d\theta_v \gets 0$
    \For{$(s, a) \in \textsc{TreeIterator}(\mathbf{s^*})$}
        \State \textbf{// Compute the future rewards for the given action}
        \State $R \gets - (c\cdot f(\textsc{Time}(s)) + (1-c)\cdot f(\textsc{Space}(s)))$
        \State \textbf{// Accumulate gradients wrt. policy gradient loss}
        \State $d\theta \gets d\theta + \nabla_\theta \log \pi(a|s; \theta)(R - V(s; \theta_v))$
        \State \textbf{// Accumulate gradients wrt. value function loss}
        \State $d\theta_v \gets d\theta_v + \partial(R - V(s;\theta_v))^2 / \partial\theta_v$

    \EndFor
    \State Perform update of $\theta$ using $d_\theta$ and $\theta_v$ using $d\theta_v$.

    \State $n \gets n + 1$
\EndWhile

\end{algorithmic}

\vspace{0.1in}
\textbf{Subroutines}:
\begin{itemize}[noitemsep,nolistsep]
    \item \textsc{Reset}(s): Reset the tree $s$ to its initial state.

    \item \textsc{GrowTreeDFS}(s, a): Apply action $a$ to tree node $s$, and return the next non-terminal leaf node in the tree in depth-first traversal order.

    \item \textsc{TreeIterator}(s): Non-terminal tree nodes of the subtree $s$ and their taken action.

    \item \textsc{Time}(s): Upper-bound on classification time to query the subtree $s$. In non-partitioned trees this is simply the depth of the tree.

    \item \textsc{Space}(s): Memory consumption of the subtree $s$.
\end{itemize}
\end{flushleft}
\label{alg:training}
\end{algorithm}

\vskip 0.1in
\noindent
{\bf Incorporating existing heuristics.}  \sysname can easily incorporate additional heuristics to improve the decision trees it learns. One example is adding rule partition actions. In addition to the cut action, in our \sysname implementation we also allow two types of partition actions:

\begin{enumerate}
\item \textbf{Simple}: the current node is partitioned along a single dimension using a learned threshold.
\item \textbf{EffiCuts}: the current node is partitioned using the EffiCuts partition heuristic \cite{EffiCuts}.
\end{enumerate}

\begin{figure}[t]
\centering
    \includegraphics[width=0.95\linewidth]{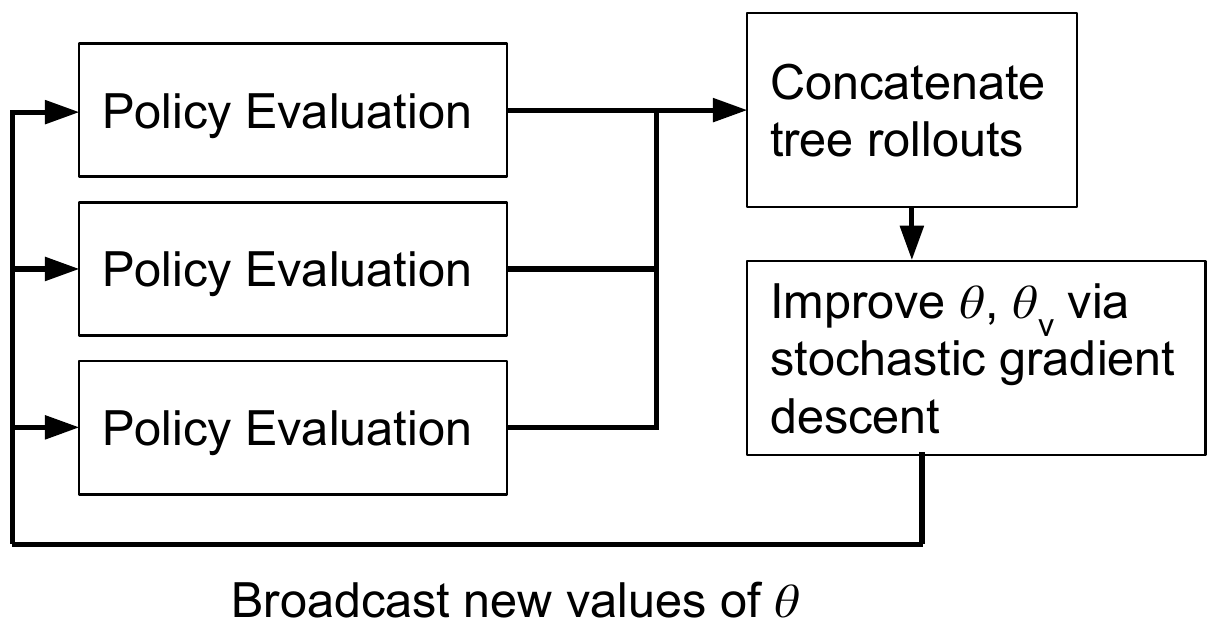}
\vspace{-0.1in}
\caption{\sysname can be parallelized by generating decision trees in
parallel from the current policy.}
\vspace{-0.1in}
\label{fig:parallel}
\end{figure}

\vskip 0.1in
\noindent
{\bf Scaling out to handle large packet classifiers.} The pseudocode in
Algorithm~\ref{alg:training} is for a single-threaded implementation of \sysname.
This is sufficient for small classifiers. But for large classifiers with tens or
hundreds of thousands of rules, parallelism can significantly improve the speed of training. In Figure \ref{fig:parallel} we show how Algorithm~\ref{alg:training} can be adapted to build multiple decision trees in parallel.

\vskip 0.1in
\noindent
{\bf Handling classifier updates.} Packet classifiers are often updated by network
operators based on application requirements, e.g., adding access control rules
for new devices.
For small updates of only a few rules, \sysname
modifies the existing decision tree to reflect the changes.
New rules are added to the
decision tree according to the existing structure; deleted rules
are removed from the terminal leaf nodes. When enough small
updates accumulate or a large update is made to the classifier, \sysname re-runs training.

\section{Implementation}
\label{sec:implementation}

Deep RL algorithms are notoriously difficult to reproduce \cite{henderson2017}. For a practical implementation, we prioritize the ability to $(i)$ leverage off-the-shelf RL algorithms, and $(ii)$ easily scale {\sysname} to enable parallel training of policies. 

\parabf{Decision tree implementation.} We implement the decision tree data structure for {\sysname} in Python for ease of development. To ensure minor implementation differences do not bias our results, we use this same data structure to implement each baseline algorithm (e.g., HiCuts, EffiCuts, etc.), as well as to implement {\sysname}.

\parabf{Branching decision process environment.} As discussed in Section \ref{sec:design}, the branching structure of the {\sysname} environment poses a challenge due to its mismatch with the MDP formulation assumed by many RL algorithms. A typical RL environment defines a transition function $P_a(s_{t+1}|s_t)$ and a reward function $R_a(s, s')$. The first difference is that the state transition function in {\sysname} returns multiple child states, instead of a single state., i.e., $(s_t, a_t) \rightarrow \{s^0_{t+1}, ..., s^k_{t+1}\}$.
Second, the final reward for {\sysname} is computed by aggregating across the rewards of child states. More precisely, for the cut action we use \texttt{max} aggregation for classification time and \texttt{sum} aggregation for memory footprint. For the partition action, we use \texttt{sum} aggregation for both metrics. 

The recursive dependence of the {\sysname} reward calculation on all descendent state actions means that it is difficult to flatten the tree structure of the environment into a single MDP, which is required by existing off-the-shelf RL algorithms. Rather than attempting to flatten the {\sysname} environment, our solution is to instead treat the {\sysname} environment as a series of \textit{independent} 1-step decision problems, each of which yields an ``immediate'' reward. The actual reward for these 1-step decisions is calculated once the relevant sub-tree rollout is complete.

For example, consider a {\sysname} tree rollout from a root node $s_1$. Based on $\pi_\theta$ the agent decides to take action $a_1$ to split $s_1$ into $s_2$, $s_3$, and $s_4$. Of these child nodes, only $s_4$ needs to be further split (via $a_2$), into $s_5$ and $s_6$, which finishes the tree. The experiences collected from this rollout consist of two independent 1-step rollouts: ($s_1$, $a_1$) and ($s_4$, $a_2$). Taking the time-space coefficient $c=1$ and discount factor $\gamma=1$ for simplicity, the total reward $R$ for each rollout would be $R=2$ and $R=1$ respectively.

\parabf{Multi-agent implementation.} Since these 1-step decisions are logically independent of each other, {\sysname} execution can be realized as a multi-agent environment, where each node's 1-decision problem is taken by an independent ``agent'' in the environment. Since we want to learn a single policy, $\pi_\theta$, for all states, the agents must be configured to share the same underlying stochastic neural network policy. This ensures all experiences go towards optimizing the single shared policy $\pi_\theta$. When using an actor critic algorithm to optimize the policies of such agents, the relevant loss calculations induced by this multi-agent realization are identical to those presented in Algorithm \ref{alg:training}.

There are several ways to implement the 1-step formulation of {\sysname} while leveraging off-the-shelf RL libraries. In Algorithm \ref{alg:training} we show standalone single-threaded pseudocode assuming a simple actor-critic algorithm is used. In our experiments, we use the multi-agent API provided by Ray RLlib \cite{rllib}, which implements parallel simulation and optimization of such RL environments.

\parabf{Performance.} We found that {\sysname} often converges to its optimal solution within just a few hundred rollouts. The size of the rule set does not significantly affect the number of rollouts needed for convergence, but affects the running time of each rollout. For smaller problems (e.g., 1000 rules), this may be within a few minutes of CPU time. The computational overhead for larger problem scales with the size of the classifier, i.e., linearly with the number of rules that must be scanned per action taken to grow the tree. The bulk of time in {\sysname} is spent executing tree cut actions. This is largely an artifact of our Python implementation, which iterates over each rule present in a node on each cut action. An optimized C++ implementation of the decision tree would further reduce the training time. 

\subsection{Optimizations}

\parabf{Rollout truncation.} During the initial phase of learning, the unoptimized policy will create excessively large trees. Since {\sysname} does not start learning until a tree is complete, it is necessary to truncate rollouts to speed up the initial phase of training. For larger classifiers, we found it necessary to allow rollouts of up to $15000$ actions in length.

\parabf{Depth truncation.} Since valid solutions never involve trees of depth greater than a few hundred, we also truncate trees once they reach a certain depth. In our experience, depth truncation is only a factor early on in learning; {\sysname} quickly learns to avoid creating very deep trees.

\parabf{Proximal Policy Optimization.} For better stability and more sample-efficient learning, in our experiments we choose to use Proximal Policy Optimization (PPO) \cite{ppo}. PPO implements an actor-critic style loss with entropy regularization and a clipped surrogate objective, which enables improved exploration and sample efficiency. We report the PPO hyperparameters we used in Appendix \ref{appendix:hyperparams}. It is important to note however that this particular choice of RL algorithm is not fundamental to {\sysname}.

\section{Evaluation}
\label{sec:evaluation}

\begin{figure*}[ht]
\centering
    \includegraphics[width=\linewidth,clip,trim={0.8cm 0.5cm 0.5cm 0.4cm}]{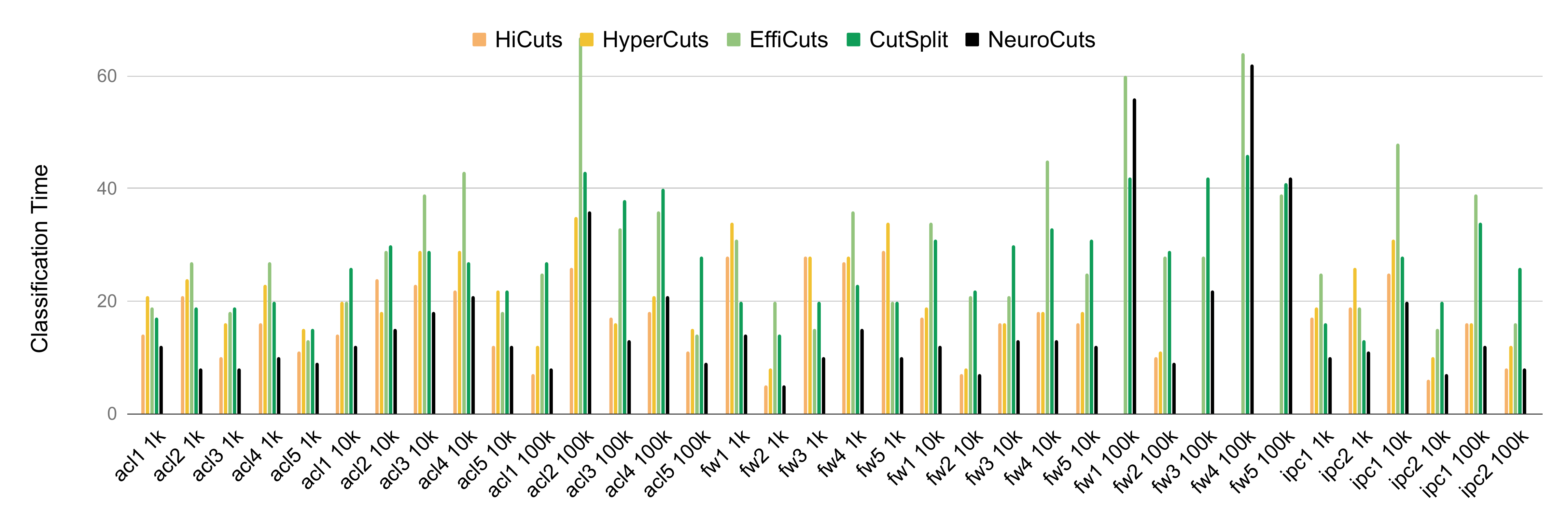}
\caption{Classification time (tree depth) for HiCuts, HyperCuts, EffiCuts, and {\sysname} (time-optimized). We omit four entries for HiCuts and HyperCuts that did not complete after more than 24 hours.}
\label{fig:time_eval}
\end{figure*}
\begin{figure*}[ht]
\centering
    \includegraphics[width=\linewidth,clip,trim={0.8cm 0.5cm 0.5cm 0.4cm}]{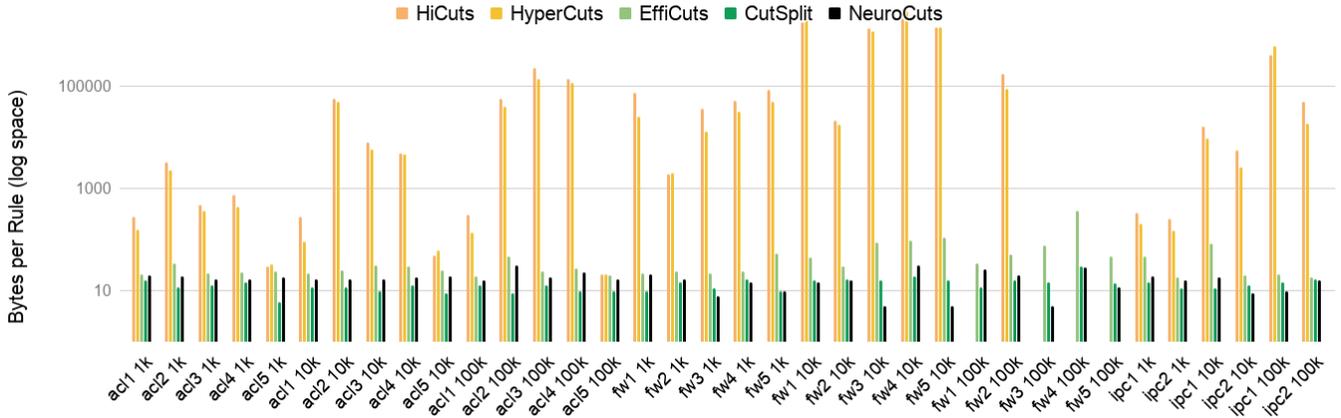}
\caption{Memory footprint (bytes per rule) used for HiCuts, HyperCuts, EffiCuts, and {\sysname} (space-optimized). We omit four entries for HiCuts and HyperCuts that did not complete after more than 24 hours.}
\label{fig:space_eval}
\end{figure*}

In the evaluation, we seek to answer the following questions:

\begin{enumerate}
    \item How does {\sysname} compare to the state-of-the-art approaches in terms of classification time and memory footprint? (Section \ref{ssec:time} and \ref{ssec:space})
    \item Beyond \textit{tabula rasa} learning, can {\sysname} effectively incorporate and improve upon pre-engineered heuristics? (Section \ref{ssec:efficuts})
    \item How much influence does the time-space coefficient $c$ have on the performance of {\sysname}? (Section \ref{ssec:tuning})
\end{enumerate}

For the results presented in the next sections, we evaluated {\sysname} using the range of hyperparameters shown in Appendix \ref{appendix:hyperparams}. We did not otherwise perform extensive hyperparameter tuning; in fact we use close to the default hyperparameter configuration of the PPO algorithm. The notable hyperparameters we swept over include:

\begin{itemize}
\item Allowed top-node partitioning (none, simple, and the EffiCuts heuristic), which strongly biases {\sysname} towards learning trees optimized for time (none) vs space (EffiCuts), or somewhere in the middle (simple).
\item The max number of timesteps allowed per rollout before truncation. It must be large enough to enable solving the problem, but not so large that it slows down the initial phase of training.
\item We also experimented with values for the time-space tradeoff coefficient $c \in \{0, 0.1, 0.5, 1\}$. When $c < 1$, we used $\textsc{log}(x)$ as the reward scaling function to simplify the combining of the time and space rewards.
\end{itemize}

We ran {\sysname} on m4.16xl AWS machines, with four CPU cores used per NeuroCuts instance to speed up the experiment. Because the neural network model and data sizes produced by {\sysname} are quite small (e.g., in contrast to image observations from Atari games), the use of GPUs is not necessary. Our main training bottleneck was the Python implementation of the decision tree. We ran each {\sysname} instance for up to 10 million timesteps (i.e., up to a couple thousand generated trees in total), or until convergence.

We compare {\sysname} with four hand-tuned algorithms: HiCuts~\cite{HiCuts}, HyperCuts~\cite{HyperCuts}, EffiCuts~\cite{EffiCuts}, and CutSplit~\cite{cutsplit}. We use the standard benchmark, ClassBench~\cite{classbench}, to generate packet classifiers with different characteristics and sizes.
We use two metrics: classification time (tree depth) and memory footprint (bytes per rule).

We find that \sysname significantly improves over all baselines in classification time while also generating significantly more compact trees. \sysname is also competitive when optimizing for memory, with a 25\% median space improvement over EffiCuts without compromising in time.

\subsection{Time-Optimized {\sysname}}
\label{ssec:time}

In Figure~\ref{fig:time_eval}, we compare the best time-optimized trees generated by {\sysname} against HiCuts, HyperCuts, EffiCuts, and CutSplit in the ClassBench classifiers. {\sysname} provides a 20\%, 38\%, 52\% and 56\% median improvement over HiCuts, HyperCuts, EffiCuts, and CutSplit respectively. {\sysname} also does better than the minimum of all baselines in 70\% of the cases, with a median all-baseline improvement of 18\%, average improvement of 12\%, and best-case improvement of 58\%. These time-optimized trees generally correspond to {\sysname} runs with either no partitioning action or the simple top-node partitioning action. 

\subsection{Space-Optimized {\sysname}}
\label{ssec:space}

We again compare {\sysname} against the baselines in Figure~\ref{fig:space_eval}, this time selecting the most space-optimized trees and comparing the memory footprint (bytes per rule). As expected, {\sysname} does significantly better than HiCuts and HyperCuts since it can learn to leverage the partition action. NeuroCut's space-optimized trees show a 40\% median and 44\% mean improvement over EffiCuts. In our experiments {\sysname} does not usually outperform CutSplit in memory footprint, with a 26\% higher median memory usage compared to CutSplit, though the best case improvement is still 3$\times$ (66\%) \textit{over all baselines}.

Separately, we also note that the memory footprints of the best \textit{time-optimized trees} generated by {\sysname} are significantly lower than those generated by HiCuts and HyperCuts, with a >100$\times$ \textit{median} space improvement along with the better classification times reported in Section \ref{ssec:time}. However, these time-optimized trees are not competitive in space with the space-optimized {\sysname}, EffiCuts and CutSplit trees.

\subsection{Improving on EffiCuts}
\label{ssec:efficuts}

\begin{figure}
  \centering
      \vspace{-.1in}
  \begin{subfigure}[{\sysname} can build on the EffiCuts partitioner to generate trees up to 10$\times$ (90\%) more space efficient than EffiCuts. In this experiment {\sysname} did as well or better than EffiCuts on all 36 rule sets.]{
      \includegraphics[width=\linewidth]{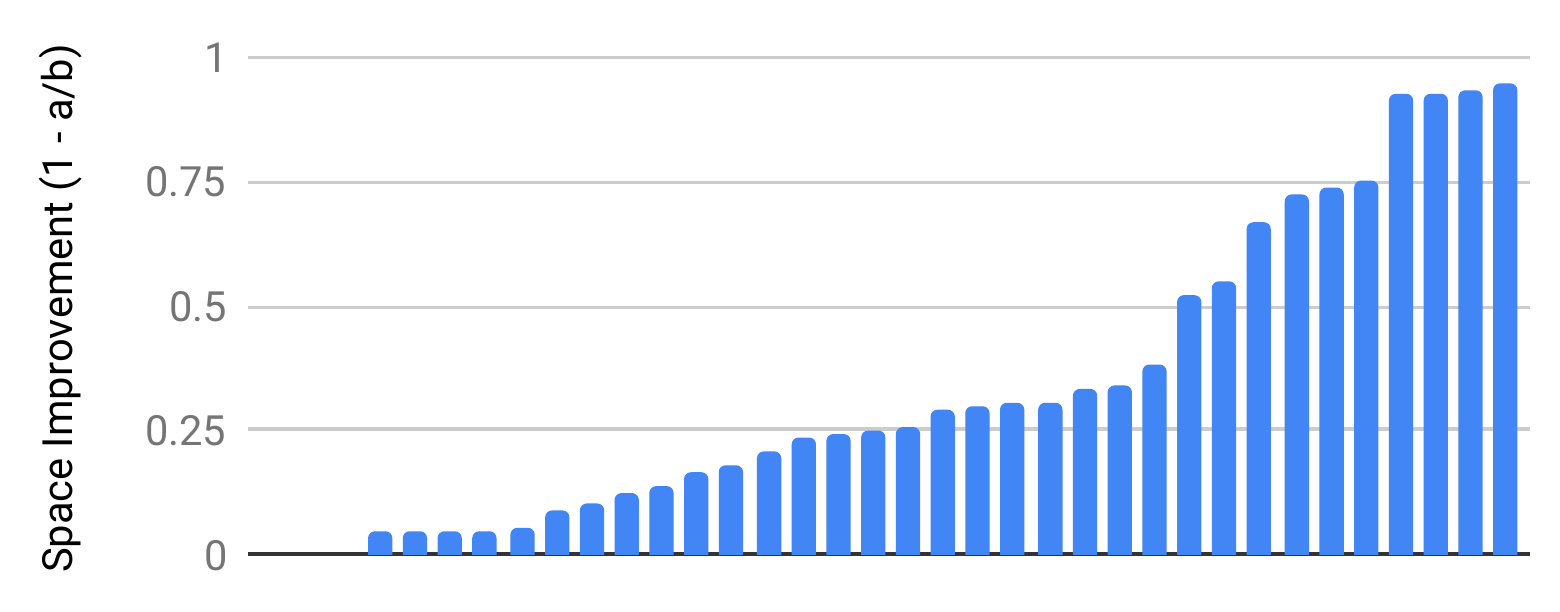}
  }
  \end{subfigure}
      \vspace{-.1in}
  \begin{subfigure}[{\sysname} with the EffiCuts partitioner generates trees with about the same time efficiency as EffiCuts.]{
    \includegraphics[width=\linewidth]{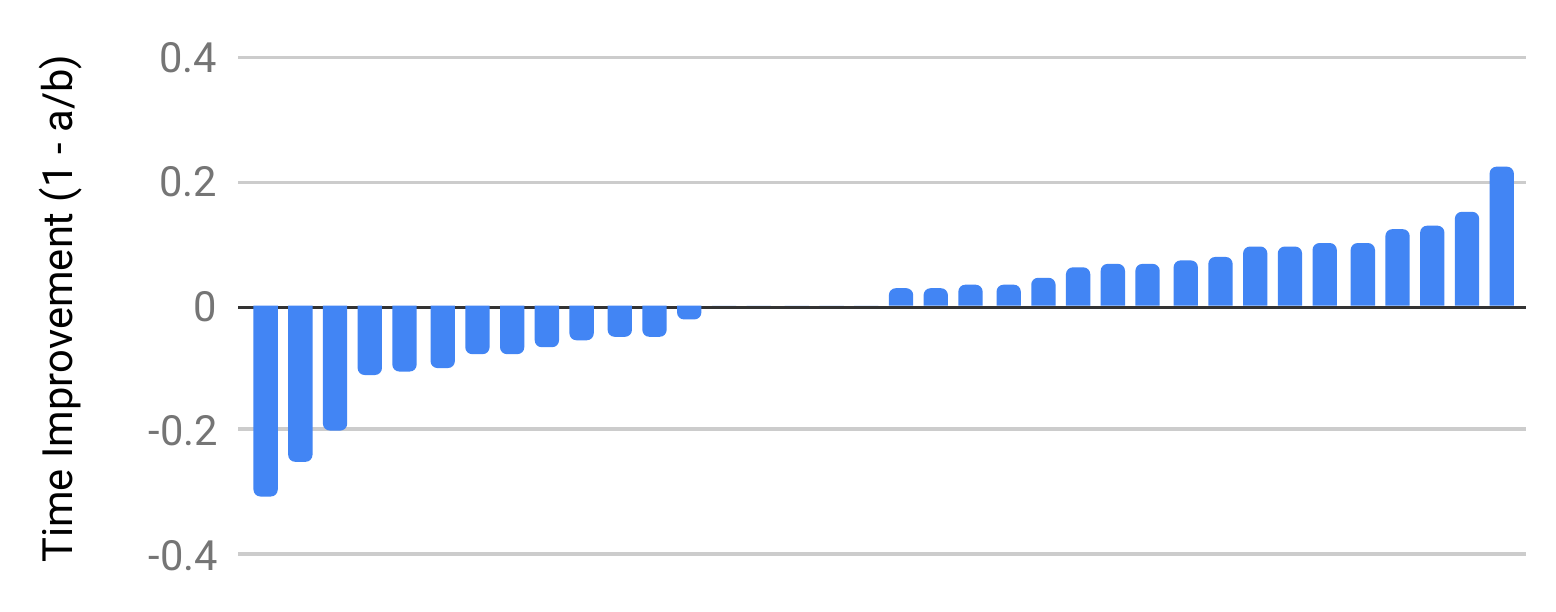}
  }
  \end{subfigure}
    \vspace{-.1in}
    \caption{Sorted rankings of {\sysname}' improvement over EffiCuts in the ClassBench benchmark. Here {\sysname} is run with only the EffiCuts partition method allowed. Positive values indicate improvements.}
    \label{fig:vs_efficuts}
    \vspace{-.05in}
\end{figure}

In Figure \ref{fig:vs_efficuts} we examine a set of 36 {\sysname} trees (one tree for each ClassBench classifier) generated by {\sysname} with the EffiCuts partition action. This is in contrast with the prior experiments that selected trees optimized for either space or time alone. On this 36-tree set, there is a median space improvement of 29\% relative to EffiCuts; median classification time is about the same. This shows that {\sysname} is able to effectively incorporate and improve on pre-engineered heuristics such as the EffiCuts top-level partition function.

Surprisingly, {\sysname} is able to outperform EffiCuts despite the fact that {\sysname} does not use multi-dimensional cut actions. When we evaluate EffiCuts with these cut types disabled, the memory advantage of {\sysname} widens to 67\% at the median. This suggests that {\sysname} could further improve its performance if we also incorporate multi-dimensional cut actions via parametric action encoding techniques \cite{fbhorizon}. It would also be interesting to, besides adding actions to {\sysname}, consider postprocessing steps such as resampling that can be used to further improve the stochastic policy output.

\subsection{Tuning Time vs Space}
\label{ssec:tuning}

\begin{figure}
  \centering
      \includegraphics[width=\linewidth]{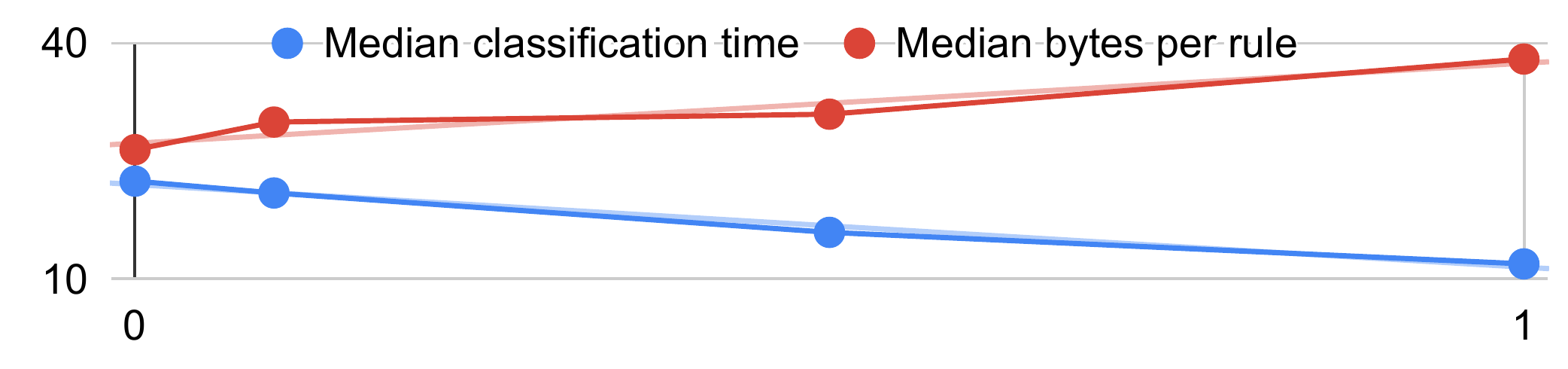}
      \small{value of $c$}
    \caption{The classification time improves by 2$\times$ as the time-space coefficient $c \rightarrow 1$, and conversely, number of bytes per rule improves 2$\times$ as $c \rightarrow 0$.}
    \label{fig:tradeoff}
\end{figure}

Finally, in Figure \ref{fig:tradeoff} we sweep across a range of values of $c$ for {\sysname} with the simple partition method and \textsc{log}(x) reward scaling. We plot the ClassBench median of the best classification times and bytes per rule found for each classifier.
We find that classification time improves by $2\times$ as $c \rightarrow 1$, while the number of bytes per rule improves 2$\times$ as $c \rightarrow 0$.
This shows that $c$ is effective in controlling the tradeoff between space and time.

\section{Related Work}
\label{sec:related}

\parabf{Packet classification.}
Packet classification is a long-standing problem in computer networking.
Decision-tree based algorithms are a major class of algorithmic solutions.
Existing solutions rely on hand-tuned heuristics to build decision trees.
HiCuts~\cite{HiCuts} is a pioneering work in this space.
It cuts the space of each node in one dimension to create multiple equal-sized subspaces to separate rules.
HyperCuts~\cite{HyperCuts} extends HiCuts by allowing cutting in multiple dimensions at each node.
HyperSplit~\cite{hypersplit} combines the advantages of rule-based space decomposition and local-optimized recursion to guarantee worst-case classification time and reduce memory footprint.
EffiCuts~\cite{EffiCuts} introduces four heuristics, including separable trees, tree merging, equal-dense cuts and node co-location, to reduce rule replication and imbalance cutting.
CutSplit~\cite{cutsplit} integrates equal-sized cutting and equal-dense cutting
to optimize decision trees. Besides decision-tree based algorithms, there are also other algorithms proposed for packet classification, such as tuple space search~\cite{tss}, RFC~\cite{rfc} and
DCFL~\cite{dcfl}. These algorithms are not as popular as decision-tree based algorithms, because they are either too slow or consume too much memory. There are also solutions that exploit specialized hardware such as TCAMs, GPUs and FPGAs to support packet classification~\cite{tcam, tcam-sigcomm12, sax-sigcomm14, tcam-sigcomm15, redund-tcam, multilayer-gpu, click-gpu, fpga-gpu}. Compared to existing work,
\sysname is an algorithmic solution that applies Deep RL to generate efficient decision trees, with the capability to incorporate and improve on existing heuristics as needed.

\parabf{Decision trees for machine learning.} There have been several proposals to use deep learning to optimize the performance of decision trees for machine learning problems \cite{norouzi2015efficient, xiong2017learning, Kontschieder_2015_ICCV}. In these settings, the objective is maximizing test accuracy. In contrast, packet classification decision trees provide perfect accuracy by construction, and the objective is minimizing classification time and memory usage.

\parabf{Structured data in deep learning.} There have many recent proposals towards applying deep learning to process and generate tree and graph data structures \cite{zhou2018graph, you2018graph, guez2018learning, graph-emb, graphemb-aaai, non-graphemb}. \sysname sidesteps the need to explicitly process graphs, instead exploiting the structure of the problem to encode agent state into a compact fixed-length representation.

\parabf{Deep reinforcement learning.}
Deep RL leverages the modeling capacity of deep neural networks to extend classical RL to domains with large, high-dimensional state and action spaces.
DQN~\cite{dqn-atari, dqn, dql} is one of the earliest successes of Deep RL, and 
shows how to learn control policies from high-dimensional sensory inputs and achieve human-level 
performance in Atari 2600 games.
A3C, PPO, and IMPALA \cite{a3c, ppo, espeholt2018impala} scale actor-critic algorithms to leverage many parallel workers.
AlphaGo~\cite{alphago}, AlphaGo Zero~\cite{alphagozero} and AlphaZero~\cite{generalalphago} show that Deep RL algorithms can achieve superhuman
performance in many challenging games like Go, chess and shogi.
Deep RL has also been applied to many other domains like
natural language processing~\cite{nlp-drl} and robotics~\cite{e2etraining, handeye-coord, robotic-grasp}.
\sysname works in a discrete environment and applies Deep RL to learn decision trees for packet classification.

\parabf{Deep learning for networking and systems.} Recently there has been an uptake in applying deep
learning to networking and systems problems~\cite{ddl, nas, pensieve, drl-route, schedule-drl, drl-cc, drl-pcc, verus, resource-drl}.
NAS~\cite{nas} utilizes client computation and deep neural networks
to improve the
video quality independent to the available bandwidth.
Pensieve~\cite{pensieve} generates adaptive bitrate algorithms using Deep RL 
without relying on pre-programmed models or assumptions about the environment.
Valadarsky \etal~\cite{drl-route}
applies Deep RL to learn network routing.
Chinchali \etal~\cite{schedule-drl} uses Deep RL for traffic scheduling in cellular networks.
AuTO~\cite{auto-drl} scales Deep RL for datacenter-scale traffic optimization.
There are also many solutions that apply deep reinforcement learning to
congestion control~\cite{drl-cc,drl-pcc,verus} and resource management~\cite{resource-drl}.
We explore the application of Deep RL to packet classification, and propose a new algorithm to learn decision trees with succinct encoding and scalable training mechanisms.

\section{Conclusion}
\label{sec:conclusion}

We present NeuroCuts, a simple and effective Deep RL formulation of the packet classification problem. NeuroCuts provides significant improvements on classification time and memory footprint compared to state-of-the-art algorithms. It can easily incorporate pre-engineered heuristics to leverage their domain knowledge, optimize for flexible objectives, and generates decision trees which are easy to test and deploy in any environment.

We hope \sysname can inspire a new generation of learning-based algorithms for packet classification. As a concrete example, \sysname currently optimizes for the worst-case classification time or memory footprint. By considering a specific traffic pattern, \sysname can be extended to other objectives such as average classification time. This would allow \sysname to not only optimize for a specific classifier but also for a specific traffic pattern in a given deployment.

\clearpage
{
\bibliographystyle{ACM-Reference-Format}
\balance
\bibliography{neurocuts}}

\appendix

\section{NeuroCuts Action and Observation Spaces}
\label{appendix:spaces}
NeuroCuts action and observation spaces described in OpenAI Gym format \cite{gym}. Actions are sampled from two categorical distributions that select the dimension and action to perform on the dimension respectively. Observations are encoded in a one-hot bit vector (278 bits in total) that describes the node ranges, partitioning info, and action mask (i.e., for prohibiting partitioning actions at lower levels).

\subsection{Action Space}
Tuple(Discrete(NumDims),\\
Discrete(NumCutActions + NumPartitionActions))

\subsection{Observation Space}
Box(low=0, high=1, shape=(278,))

\subsection{Observation Components}
 (BinaryString($Range_{min}^{dim}$) + BinaryString($Range_{max}^{dim}$) +\\
 OneHot($Partition_{min}^{dim}$) + OneHot($Partition_{max}^{dim}$))\\
 $\forall dim \in \{SrcIP, DstIP, SrcPort, DstPort, Protocol\}$ +\\ OneHot(EffiCutsPartitionID) + ActionMask
 
 When not using the EffiCuts partitioner, the $Partition^{dim}$ rule dimension coverage thresholds are set to one of the following discrete levels: 0\%, 2\%, 4\%, 8\%, 16\%, 32\%, 64\%, and 100\%.
 
 We note that the set of rules for the packet classifier are \textit{not} present in the observation space. \sysname learns to account for packet classifier rules implicitly through the rewards it gets from the environment.

\section{NeuroCuts Hyperparameters}
\label{appendix:hyperparams}
\begin{table}[h]
\begin{center}
\begin{tabular}{l | l}
{\bf Hyperparameter} & {\bf Value} \\\hline
Time-space coefficient $c$ & <set by user>\\
Top-node partitioning & \{none, simple, EffiCuts\}\\
Reward scaling function $f$ & \{x, \textsc{log}(x)\}\\
Max timesteps per rollout & \{1000, 5000, 15000\}\\
Max tree depth & \{100, 500\}\\
Max timesteps to train & 10000000\\
Max timesteps per batch & 60000\\
Model type & fully-connected\\
Model nonlinearity & tanh\\
Model hidden layers & [512, 512]\\
Weight sharing between $\theta, \theta_v$ & true\\
Learning rate & 0.00005\\
Discount factor $\gamma$ & 1.0\\
PPO entropy coefficient & 0.01\\
PPO clip param & 0.3\\
PPO VF clip param & 10.0\\
PPO KL target & 0.01\\
SGD iterations per batch & 30\\
SGD minibatch size & 1000\\
\end{tabular}
\end{center}
\caption{NeuroCuts hyperparameters. Values in curly braces denote a set of values searched over during evaluation. We found that the most sensitive hyperparameter is the top-node partitioning, which greatly affects the structure of the search problem. It is also important to ensure that the rollout timestep limit and model used are sufficiently large for the problem (we found that using 256-unit hidden layers slightly degraded learning for larger classifiers, and more severely so at 64-units).}
\label{table:hyperparams}
\end{table}

\end{document}